\def\MyApJ#1{#1}
\def\MyMNRAS#1{}

\documentclass[numberedappendix]{emulateapj}

\usepackage{graphicx}% Include figure files
\usepackage{bm}% bold math
\usepackage{epsf}
\usepackage{verbatim}
\usepackage{amsmath}
\usepackage{amsfonts}
\usepackage{epsfig}
\usepackage{dcolumn}
\usepackage{graphicx}
\usepackage{amssymb}
\usepackage{tablefootnote}

\usepackage{xcolor}
\definecolor{darkgreen}{rgb}{0.0,0.5,0.0}
\usepackage[colorlinks,citecolor=darkgreen]{hyperref} % Works with PDFLaTeX

\usepackage[percent]{overpic}

\def\myfig#1{./#1}

\def\PlotFigs#1{} % for fast, figure-free compilation
\MyApJ{\bibliographystyle{apj}}
\MyMNRAS{\bibliographystyle{mnras}}

\defcitealias{KeshetEtAl17}{K17}
\newcommand{\Coma}{{\citetalias{KeshetEtAl17}}}

\defcitealias{ReissKeshet18}{R18}
\newcommand{\Stack}{{\citetalias{ReissKeshet18}}}

\defcitealias{KeshetReiss18}{KR18}
\newcommand{\ComaB}{{\citetalias{KeshetReiss18}}}

\defcitealias{HurierEtAl19}{H19}
\newcommand{\SZA}{{\citetalias{HurierEtAl19}}}

\newcommand{\ie}{\emph{i.e.,} }

\newcommand{\be}{\begin{equation}}
\newcommand{\ee}{\end{equation}}
\newcommand{\bea}{\begin{equation*}}
\newcommand{\eea}{\end{equation*}}
\newcommand{\beqr}{\begin{eqnarray} \nonumber}
\newcommand{\eeqr}{\end{eqnarray}}
\newcommand{\beqrb}{\begin{eqnarray}}
\newcommand{\eeqrb}{\nonumber \end{eqnarray}}
\newcommand{\fin}{\mbox{ .}}
\newcommand{\coma}{\mbox{ ,}}

\newcommand{\se}{\mbox{ s}}
\newcommand{\yr}{\mbox{ yr}}

\newcommand{\km}{\mbox{ km}}

\newcommand{\kpc}{\mbox{ kpc}}
\newcommand{\Mpc}{\mbox{ Mpc}}

\newcommand{\keV}{\mbox{ keV}}

\newcommand{\GeV}{\mbox{ GeV}}
\newcommand{\TeV}{\mbox{ TeV}}

\newcommand{\const}{\mbox{const.}}

\newcommand{\gama}{$\gamma$}

\newcommand{\vect}[1]{\bm{#1}}
\newcommand{\dgr}{{^\circ}}
\newcommand{\till}{{\mbox{--}}}

\newcommand{\mass}{\bar{m}}

\newcommand{\myTS}{\mbox{TS}}

\newcommand{\model}{\mu}
\newcommand{\Mbin}{\mathcal{M}}

\def\mybm#1{#1}

\MyApJ{
\begin{document}
\title{Coincident SZ and $\gamma$-ray signals from cluster virial shocks}
\shorttitle{Coincident SZ and gamma-ray virial signals}
\author{Uri Keshet$^{1,\dagger}$}
\author{Ido Reiss$^{2}$}
\author{Guillaume Hurier$^{3}$}
\affil{$^{1}$Physics Department, Ben-Gurion University of the Negev, POB 653, Be'er-Sheva 84105, Israel}
\affil{$^{2}$Department of Physics, NRCN, POB 9001, Be'er Sheva 84190, Israel}
\affil{$^{3}$Centro de Estudios de Fisica del Cosmos de Aragon, Plaza San Juan 1, Planta-2, 44001, Teruel, Spain}
\thanks{$^{\dagger}$Electronic address: ukeshet@bgu.ac.il}
\shortauthors{Keshet et al.}
\date{\today}
}

\MyMNRAS{
\title[Coincident SZ and $\gamma$-ray virial signals]{Coincident SZ and $\gamma$-ray signals from cluster virial shocks}
\author[Keshet et al.]{
Uri Keshet$^{1}$\thanks{E-mail: ukeshet@bgu.ac.il},
Ido Reiss$^{1,2}$,
and Guillaume Hurier$^{3}$
\\
% List of institutions
$^{1}$Physics Department, Ben-Gurion University of the Negev, POB 653, Be'er Sheva 84105, Israel\\
$^{2}$Department of Physics, NRCN, POB 9001, Be'er Sheva 84190, Israel\\
$^{3}$Centro de Estudios de Fisica del Cosmos de Aragon, Plaza San Juan 1, Planta-2, 44001, Teruel, Spain
}
\pubyear{2017}
\begin{document}
\label{firstpage}
\pagerange{\pageref{firstpage}--\pageref{lastpage}}
\maketitle
}

\begin{abstract}
Virial shocks around galaxy clusters are expected to show a cutoff in the thermal Sunyaev-Zel'dovich (SZ) signal, coincident with a leptonic ring.
However, until now, leptonic virial signals were reported only in Coma and in stacked \emph{Fermi}-LAT clusters, and an SZ virial shock signal was reported only in A2319.
We point out that a few clusters --- presently Coma, A2319, and A2142 --- already show a sharp drop in \emph{Planck} SZ pressure near the virial radius, coincident with a weak LAT $\gamma$-ray excess.
These signatures are shown to trace the virial shocks of the clusters, at joint medium to high confidence levels.
The electron acceleration rates inferred from $\gamma$-rays are consistent with previous measurements.
The combined signal allows a separate measurement of the $\sim0.5\%$ acceleration efficiency and of the accretion rate.
Lower limits of order a few are imposed on the shock Mach numbers.
\end{abstract}

\MyMNRAS{
}

\MyApJ{
%\keywords{galaxies: clusters: individual (Coma, A2319, A2142) --- gamma rays: galaxies: clusters ---  Galaxies: clusters: intracluster medium --- acceleration of particles -- shock waves}}
\maketitle
}

\section{Introduction}
\label{sec:Intro}

As a galaxy cluster grows, by accreting matter from its surroundings, a strong, collisionless shock is thought to form and heat the inflow to virial temperatures, at the so-called virial shock radius, $r_s$. In spite of considerable efforts, these elusive shocks have only recently been traced, thanks to their distinct leptonic signature.
However, these signals have not yet been corroborated by independent, or more direct, indicators.

By analogy with supernova remnant (SNR) shocks, virial shocks too should accelerate charged particles to highly relativistic, $\gtrsim 10\TeV$ energies.
These particles, known as cosmic ray (CR) electrons (CREs) and ions (CRIs), should show a nearly flat, \ie $E^2dN/dE\simeq \const$, spectrum (equal energy per logarithmic CR energy bin), radiating a distinctive non-thermal signature which stands out at the extreme ends of the electromagnetic spectrum.
High-energy CREs cool rapidly, on timescales much shorter than the Hubble time $H^{-1}$, mainly by Compton-scattering cosmic microwave-background (CMB) photons
\citep{LoebWaxman00, TotaniKitayama00, KeshetEtAl03}.
These up-scattered photons should then produce \gama-ray emission in a thin shell around the galaxy cluster, as anticipated analytically \citep{WaxmanLoeb00, TotaniKitayama00} and calibrated using cosmological simulations \citep{KeshetEtAl03, Miniati02}.
The projected \gama-ray signal typically shows an elliptic morphology, elongated towards the large-scale filaments feeding the cluster \citep{KeshetEtAl03,KeshetEtAl04}.
The same \gama-ray emitting CREs are also expected to generate an inverse-Compton ring in the optical band \citep{YamazakiLoeb15} and in hard X-rays \citep{KushnirWaxman10}, and a synchrotron ring in radio frequencies \citep{WaxmanLoeb00, KeshetEtAl04, KeshetEtAl04_SKA}.
It was recently shown that the shocks can also be detected in soft X-rays, below the peak energy of the thermal component \citep[][henceforth \ComaB]{KeshetReiss18}.

By stacking \emph{Fermi} Large Area Telescope (LAT; henceforth) data around 112 massive clusters, the cumulative \gama-ray emission from multiple virial shocks was detected at a high ($\sim5.8\sigma$) significance \citep[][henceforth \Stack]{ReissEtAl17, ReissKeshet18}.
The signal peaks upon radial binning around the expected shock radius, $r_s\simeq(2.3\pm0.1)R_{500}\simeq 1.5R_{200}$, indicating a flat CRE spectral index, $p\equiv -d\ln N/d\ln E=2.1\pm0.2$.
Here, subscripts $\delta=200$ or $500$ designate an enclosed mass density that is $\delta$ times larger than the critical mass density of the Universe.
This signal indicates that the stacked shocks deposit on average $\xi_e\dot{m}=(0.6\pm0.1)\%$ (with a systematic uncertainty factor of $\sim 2$) of the thermal energy in CREs over a Hubble time.
Here, $\xi_e$ is the fraction of shocked thermal energy deposited in CREs, and $\dot{m}\equiv \dot{M}/(M H)$ is a dimensionless accretion rate of order a few, where $M$ is the cluster mass and $H$ is the Hubble parameter.
As these results were obtained by radial binning, they sample only the radial component of the projected virial shocks, somewhat diluting the shock signal.

It is interesting to study the signal from individual nearby clusters, where the signal may be picked up directly, without stacking.
The Coma cluster (Abell 1656), in particular, is one of the richest nearby clusters, and is exceptionally suitable for the search for virial shock signatures.
An analysis \citep[][henceforth \Coma]{KeshetEtAl17} of a $\sim220\GeV$ VERITAS mosaic of Coma \citep{VERITAS12_Coma} found evidence for a large-scale, extended \gama-ray feature surrounding the cluster.
It is challenging to uncover the signal at lower energies, where the Galactic foreground becomes strong;
LAT studies thus imposed upper limits on various emission morphologies \citep{ZandanelAndo14, Prokhorov14, FermiComa16}.
However, the effectively higher resolution of LAT observations motivates a template similar to, but thinner than, the VERITAS signal, giving a $3.4\sigma$ LAT excess; a soft X-ray signal anticipated from lower energy CREs advected to smaller radii was also detected ($>5\sigma$).
Both signals are best fit by the VERITAS ring morphology; the VERITAS, \emph{Fermi} and \emph{ROSAT} signals all agree (within systematics) with the same CRE distribution, provided that $p\simeq 2.0\text{--}2.2$ (\ComaB).

A more direct tracer of a virial shock, independent of particle acceleration, is its imprint on the thermal Sunyaev-Zel'dovich \citep[SZ;][]{SunyaevZeldovich72} signal.
This distortion of the CMB field, produced as CMB photons traverse the hot intracluster medium (ICM), provides a direct measure of the Comptonization $y$ parameter.
The shock should then present as a sharp outward radial drop in the $y$-parameter, localized near the virial shock \citep{KocsisEtAl05}.
Preliminary evidence was found for a correlation between the \gama-ray signature in Coma and the $y$-parameter drop inferred from WMAP (\Coma).
Accurately measuring the SZ effect at sufficiently high sensitivity and resolution is challenging, but has recently become feasible thanks to the $y$-parameter maps prepared by the \emph{Planck} collaboration \citep{PlanckXXII16}.
Indeed, the first firm detection ($8.6\sigma$) of the virial SZ drop was recently reported by \citet[][henceforth \SZA]{HurierEtAl19}.
An SZ detection is clean, in the sense that a sufficiently sharp $y(r)$ drop can be unambiguously identified as an accretion shock.

Here, we present a combined analysis of \emph{Planck} $y$-parameter maps and of \emph{Fermi}-LAT data around select galaxy clusters, to test if virial shocks can be identified at a high confidence level through coinciding SZ drops and \gama-ray signals.
While photon statistics and resolution limit the LAT virial signal of an individual cluster to low confidence levels, and the drop in $y(r$) is smoothed by projection, combining \gama-rays and SZ can overcome these difficulties.
As we show, a high-significance joint detection not only supports the viability of the \gama-ray and SZ signals, but also corroborates the association of the \gama-ray signal with the virial shock, confirms that the virial shock is a strong, collisionless shock, and provides measures of the shock Mach number, the accretion rate, and the CRE acceleration efficiency.
We choose to study two clusters that already have published radial, azimuthally-averaged SZ profiles, namely Coma \citep{KhatriGaspari16} and Abell 2319 \citep{GhirardiniEtAl17}.
A third cluster --- Abell 2142 --- is selected as a test-case, based on its high mass and data availability.

The paper is arranged as follows.
Our analysis methods are presented in \S\ref{sec:Method}.
The Coma cluster is analyzed in \S\ref{sec:Coma}, A2319 in \S\ref{sec:A2319}, and A2142 in \S\ref{sec:A2142}.
The results are then summarized and discussed in \S\ref{sec:Discussion}.
We adopt a flat $\Lambda$CDM cosmological model with a Hubble constant $H_0=70\km\se^{-1}\Mpc^{-1}$ and a mass fraction $\Omega_m=0.3$.
Assuming a $76\%$ hydrogen mass fraction gives a mean particle mass $\mass\simeq 0.59m_p$.
An adiabatic index $\Gamma=5/3$ is assumed.
Confidence intervals are $68\%$ for one parameter.
The results are primarily quantified in terms of an overdensity $\delta=500$.
Accordingly, we define a normalized angular distance $\tau\equiv \theta/\theta_{500}$ from the center (defined as the X-ray peak) of the cluster.

\section{Method}
\label{sec:Method}

We extract the parameters of the analyzed clusters from the Meta Catalog of X-ray Clusters \citep[MCXC;][]{PiffarettiEtAl11}.
In addition to the location of each cluster on the sky and its $R_{500}$ radius, the catalog specifies the redshift $z$ of each cluster, so the corresponding angular radius $\theta_{500}$ can be computed.
To model the virial shock and infer the acceleration efficiency from leptonic signals, we also require (see \Stack) $M_{500}$, the mass enclosed inside $R_{500}$.
In most of the analysis, we assume that the gas distribution in each cluster is spherical.
For Coma, which shows evidence for an elongated signature at large radii, we also examine an underlying prolate distribution.
The parameters of the clusters and the results of their analyses are summarized in Table \ref{tab:Summary}.

To model the signals and estimate their significance, we use a maximal likelihood (minimal $\chi^2$) analysis.
The likelihood $\mathcal{L}$ is related to the $\chi^2$ distribution of squared normalized errors by
\begin{equation}
\label{eq:Likelihood}\ln\mathcal{L} = -\frac{1}{2}\sum_{\epsilon,\tau}\chi^2(\epsilon,\tau) \coma
\end{equation}
where the sum is carried over the photon energy bands $\epsilon$ and the radial bins $\tau$.
The test statistics \citep{MattoxEtAl96_TS} TS, defined as
\begin{equation}
\myTS \equiv -2\ln\frac{\mathcal{L}_{max,-}}{\mathcal{L}_{max,+}}=\chi^2_{-} - \chi^2_{+} \coma
\end{equation}
can then be computed.
Here, subscript '$-$' (subscript '$+$') refers to the likelihood without (with) the modelled signal, maximized over any free parameters.
Confidence levels are computed by assuming that $\myTS$ has a $\chi_\mathsf{n}^2$ distribution, where $\mathsf{n}=\mathsf{n}(+)-\mathsf{n}(-)$ is the number of free parameters added by modeling the signal \citep{Wilks1938}.

\subsection{SZ analysis}
\label{sec:MethodSZ}

We generate the radially binned profile of the Comptonization parameter in each cluster, as a vector $\bm{y}$, following the standard SZ analysis protocol \citep{PlanckInt5_2013, HurierEtAl13, PlanckXXII16, GhirardiniEtAl17}, as described in {\SZA}.
The Comptonization parameter for each line of sight $l$ is defined as
\begin{equation} \label{eq:yDef}
y=\frac{\sigma_T}{m_ec^2}\int P\,dl \coma
\end{equation}
giving the dimensionless line-of-sight integral of the electron pressure $P$.
Here, $\sigma_T$ is the Thomson cross section, $m_e$ is the electron mass, and $c$ is the speed of light.

We construct an SZ map for each cluster, with a $7\arcmin$ FWHM angular resolution, using the Modified Internal Linear Combination Algorithm \citep[MILCA;][]{HurierEtAl13}.
Each map is azimuthally averaged and binned onto concentric annuli of $2\arcmin$ thickness.
The resulting profiles are shown for Coma in Figure \ref{fig:ComaFit1}, for A2319 in Figure \ref{fig:A2319Fit1}, and for A2142 in Figure \ref{fig:A2142Fit1}.
The local background offset $y_b$ is assumed uniform in each cluster, and treated as a free parameter; the figure ordinates are offset for presentation purposes.

Due to the moderate angular resolution of the \emph{Planck} survey, the $y$-map is over-sampled, introducing correlations between pixels that propagate into the radial profiles; additional correlations are induced by the intrinsic \emph{Planck} noise properties.
The covariance matrix $C_p$ of the binned $\bm{y}$ profile is estimated using 1000 simulations of inhomogeneous correlated gaussian noise.
The error bars in the figures represent only the (square root of the) diagonal of this covariance matrix.

The $y$-map in each cluster is modelled by the line of sight integration Eq.~(\ref{eq:yDef}) over the ICM pressure.
In general, we assume spherical symmetry, $P(\vect{r})=P(r)$; deviations from sphericity are examined in \S\ref{sec:Coma}.
We first model the gas without a virial shock, using the generalized NFW profile \citep[gNFW;][]{NavarroEtAl96, NFW97},
\begin{equation} \label{eq:gNFW}
P_0(r) \propto r^{-\gamma}\left[1+C r^\alpha\right]^{(\gamma-\beta)/\alpha} \coma
\end{equation}
where $\alpha$, $\beta$, $\gamma$, and $C$ are free parameters.
We also consider a simpler, isothermal $\beta$-model,
\begin{equation} \label{eq:beta}
P_0(r) \propto \left(1+{r^2}/{r_c^2}\right)^{-3\tilde{\beta}/2} \coma
\end{equation}
where $\tilde{\beta}$ and the core radius $r_c$ are free parameters; this simpler form allows the integration to be carried out analytically.
An overall normalization and $y_b$ are two additional free parameters in each model.

The resulting $y$-map of each model is analyzed similarly to the data: convolved with a $7\arcmin$ FWHM filter and binned onto the same $2\arcmin$ radial bins, to give a binned radial vector $\bm{y}_m$.
The free parameters of the model are then determined by maximizing the likelihood.
The uncertainties are assumed to follow Gaussian statistics, such that
\begin{equation}
\chi^2 = (\bm{y}-\bm{y}_m) C_p^{-1} (\bm{y}-\bm{y}_m)^T \fin
\end{equation}

Next, each model is generalized to account for the presence of a shock.
A simple model for an internal shock is given by
\begin{equation} \label{eq:ICMShock}
P(r) = P_0(r) \times \begin{cases}
1 & \mbox{for } r\leq r_s \, ; \\
q^{-1} & \mbox{for } r>r_s \, .
\end{cases}
\end{equation}
The shock radius $r_s$ and the fractional pressure jump $q>1$ across it constitute two additional free parameters.
This model, which assumes the same pressure slope both upstream and downstream of the shock, is more appropriate for a weak ICM shock than it is for a virial shock.

For better modeling of the virial shock, we replace the upstream region $r>r_s$ with pristine accreted gas.
The infalling gas is approximated as in free fall, $v\propto r^{-1/2}$, so mass conservation implies a $\rho\propto \dot{M}/(r^2v)\propto r^{-3/2}$ mass density profile.
Adiabaticity then yields $P(r>r_s)\propto \rho^{\Gamma}\propto r^{-3\Gamma/2} = r^{-5/2}$, so the pressure profile is given by
\begin{equation} \label{eq:VirialShock}
P(r) = \begin{cases}
P_0(r) & \mbox{for } r\leq r_s \, ; \\
q^{-1}P_0(r_s) \left({r}/{r_s}\right)^{-5/2} & \mbox{for } r>r_s \, .
\end{cases}
\end{equation}
As virial shocks are expected to be strong, we also consider a model in which $q\to\infty$ in Eqs.~(\ref{eq:ICMShock}) or (\ref{eq:VirialShock}); this leaves $r_s$ as the single shock parameter.

We simultaneously fit the free parameters of the different models, with and without a shock.
In all cases, we find a substantial, at least $3\sigma$ improvement in the fit when including a virial shock.
The ICM shock profile Eq.~(\ref{eq:ICMShock}) is marginally disfavored in all cases with respect to the virial shock profile Eq.~(\ref{eq:VirialShock}).
A strong shock is favored; in some cases $q^{-1}$ is consistent with zero.
Given $q$, the Mach number $\Upsilon$ of the shock may be computed as
\begin{equation}
\Upsilon = \sqrt{\frac{q(\Gamma+1)+\Gamma-1}{2\Gamma}} = \sqrt{\frac{4q+1}{5}} \fin
\end{equation}
This relation gives an estimate or a lower limit on $\Upsilon$.

While the $\beta$ model is only a simple, special case of the gNFW model, the two models yield very similar shock parameters, well within statistical confidence intervals.
The $\beta$ model has the advantage of allowing for faster computations, both due to its fewer free parameters, and to its analytically-tractable integration along the line of sight even when introducing a shock with Eqs.~(\ref{eq:ICMShock}) or (\ref{eq:VirialShock}).
Generalizations for multiple shocks and deviations from sphericity are thus examined using the $\beta$ model.

The confidence levels of shock detection obtained in the gNFW and $\beta$ model variants are either comparable to each other, or higher in the $\beta$ model.
It is not a priori clear which model better captures the presence of the shock: the two additional free parameters of the gNFW model can follow the gas distribution better, but they may also mask the presence of the shock to some degree.
When incorporating the virial shock, the $\beta$ models give $\beta\simeq 1$, as expected at large radii, and provide a very good fit to the data; the difference in TS with respect to the gNFW+shock counterpart is $\lesssim 3$.

Several convergence tests are used to test the robustness of our results.
We examine if the results are sensitive to the radial range used in the analysis, ruling out spurious effects induced by structure at both small and large radii.
We obtain comparable results when fitting profiles in linear vs. logarithmic $r\till y$ space.
We test $y$ parameter profiles prepared with the Needlet Internal Linear Combination algorithm \citep[NILC;][]{DelabrouilleEtAl09} instead of MILCA, obtaining similar results.
Other convergence tests and variations in the models are described below, on a cluster-by-cluster basis.

\subsection{Gamma-ray analysis}
\label{sec:MethodLAT}

The \gama-ray analysis is similar to that employed in {\Stack} and in {\ComaB}.
We use the archival, $\sim8$ year, Pass-8 LAT data from the Fermi Science Support Center (FSSC)\footnote{\texttt{http://fermi.gsfc.nasa.gov/ssc}}, and the Fermi Science Tools (version \texttt{v10r0p5}).
Pre-generated weekly all-sky files are used, spanning weeks $9\till422$ for a total of $414$ weeks ($7.9\yr$), with ULTRACLEANVETO class photon events.
We consider four logarithmically-spaced energy bands in the range 1--100 GeV.
The data is discretized using a HEALPix scheme \citep{GorskiEtAl05} of order 10.
Point source contamination is minimized by masking the $95\%$ ($90\%$ for A2319; see \S\ref{sec:A2319}) containment area of each point source in the LAT 4-year point source catalog \citep[3FGL;][]{FermiPSC}.
The foreground is estimated by fitting a polynomial function to the cluster and its vicinity.

We bin the LAT data into concentric rings about the X-ray center of the cluster.
For each photon energy band $\epsilon$, and each radial bin centered on $\tau$ with width $\Delta \tau$, we define the excess emission $\Delta n\equiv n-f$ as the difference between the number $n$ of detected photons, and the number $f$ of photons estimated from the fitted foreground.
The significance of the excess emission in a given energy band $\epsilon$ and radial bin $\tau$ can then be estimated, assuming Poisson statistics with $f\gg1$, as
\begin{equation} \label{eq:SingleBinSignificance}
\nu_{\sigma}(\epsilon,\tau) \simeq {\Delta n}/{\sqrt{f}} \fin
\end{equation}

Next, we compute the $\chi^2$ contribution of the excess counts $\Delta n(\epsilon,\tau)$ with respect to the model prediction $\model(\epsilon,\tau)$, for given $\epsilon$ band and $\tau$ bin,
\begin{equation} \label{eq:ChiSquared}
\chi^2(\epsilon,\tau,\Mbin)=\frac{\left(\Delta n - \model\right)^2}{f + \model} \fin
\end{equation}
The likelihood $\mathcal{L}$ is then related to the sum over all spatial bins and energy bands, as
\begin{equation}
\label{eq:LikelihoodSum}\ln\mathcal{L} = -\frac{1}{2}\sum_{\epsilon,\tau}\chi^2(\epsilon,\tau) \fin
\end{equation}

The (so-called shell) model $\mu$ is based on leptonic emission from a thin shell in a $\beta$-model, as described in \Coma, \Stack, and \ComaB.
The two free parameters describing the shock emission are its (normalized) radius $\tau_s$, and the CRE acceleration rate $\xi_e\dot{m}$.
Another (so called planar) model we consider assumes that accretion is confined to the plane of the sky, so the emission takes the form of a ring; this model uses the same two free parameters (\ComaB).
We also consider a one-parameter model, in which the parameter $\xi_e\dot{m}=0.6\%$ is fixed on the mean value inferred from the stacking of LAT clusters (\Stack).

Such \gama-ray analyses were tested and calibrated in {\Stack} using large control catalogs, with mock clusters redistributed on the sky.
Convergence tests for all analysis parameters were carried out using a sample of 112 clusters (including A2319 and A2142) and a large mock sample.
In particular, parameters pertaining to the discretization resolution, point source removal, and foreground modeling, were shown to be well behaved.

In the above method, we analyze the \gama-ray data around A2319 and around A2142 (the \gama-ray signal from Coma was analyzed in {\ComaB}).
The radial profiles of the excess significance are presented in Figures \ref{fig:A2319_significance} (A2319) and \ref{fig:A2142_significance} (A2142).
In both clusters, we find a \gama-ray excess in the vicinity of the virial radius, and in close proximity to the shock location inferred from SZ.
We then repeat the analysis when using the high precision localisation of the virial shock by the SZ data, as a prior for the \gama-ray analysis.
Joint likelihood analyses are also carried out.

\section{Coma}
\label{sec:Coma}

% Early placement, so figure appears on the top of the next page.
\begin{figure*}
	\centerline{
        \includegraphics[trim={0 1.4cm 2.1cm 0}, clip ,width=9cm]{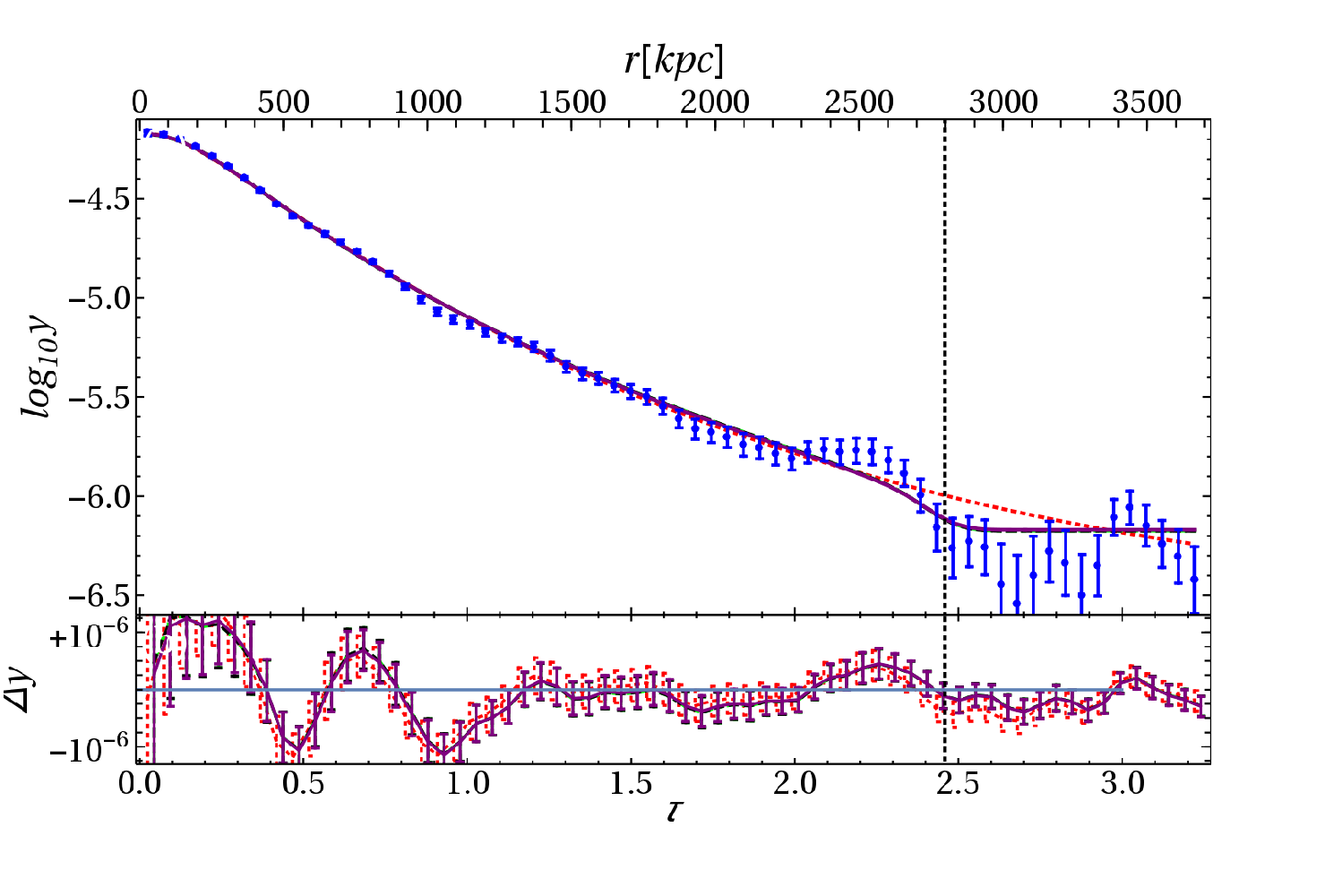}
        \includegraphics[trim={0 1.4cm 2.1cm 0}, clip ,width=9cm]{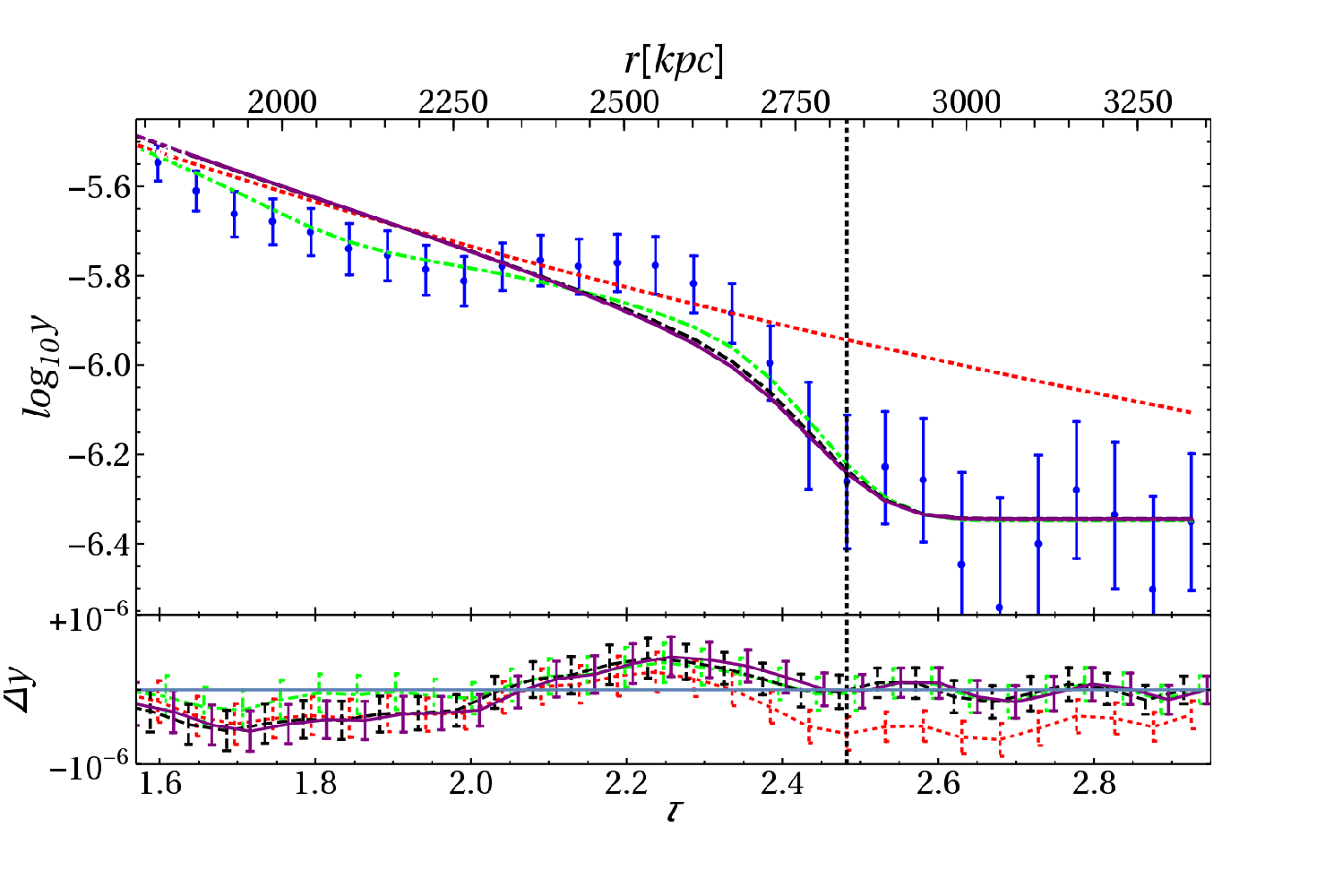}
    }
    \caption{\label{fig:ComaFit1}
    Azimuthally-averaged radial profile of the $y$-parameter in Coma, measured with \emph{Planck} ($1\sigma$ error bars) and modeled both without (dotted red curve) and with (other curves) a virial shock, in particular a strong, Mach $\Upsilon\to\infty$ spherical virial shock (purple solid curves; a dotted vertical curve shows the best-fit shock location).
    Models are based both on spherical gNFW (left panels) and isothermal $\beta$ (right, zoomed in panels) profiles; the bottom panels show the fit residuals (slightly shifted $\tau$, for visibility, in shock models).
    The left panels include models for an ICM shock (dot-dashed green) and a virial shock (dashed black) of finite $\Upsilon$; but as the inferred shock is strong, the different shock curves largely overlap.
    The right panels also include models with sharp transitions from spherical to filamentary (dot-dashed green) and to prolate (dashed black) distribution.
    }
\end{figure*}

With mass $M_{500}\simeq 4.3\times 10^{14}M_\odot$, temperature $k_B T\sim 8\keV$, richness class 2, and dimensions $R_{500}\simeq 1.1\Mpc$ and $\theta_{500}\simeq0\dgr.678$ \citep{PiffarettiEtAl11}, the Coma cluster lies only $\sim 100\Mpc$ away \citep{GavazziEtAl09}, at a redshift $z\simeq 0.0231$.
The cluster resides near the north Galactic pole (latitude $\bar{b}\sim88\dgr$), in a sky patch remarkably low on Galactic foreground, rendering it an attractive target for \gama-ray studies.
The virial radius of Coma, often defined as $R_v\simeq R_{200}\simeq 2.3\Mpc$ \citep{ReiprichBohringer02}\footnote{A wide range of $R_{200}$ estimates for Coma may be found in the literature, ranging from $1.8\Mpc$  \citep[self-similar extrapolation from the $R_{500}$ of][]{PiffarettiEtAl11}, to $2.1\Mpc$ \citep{GellerEtAl99}, $2.6\Mpc$ \citep{BrilenkovEtAl15}, and $2.8\Mpc$ \citep{KuboEtAl07}.}, corresponds to an angular radius $\theta\simeq \theta_{200}\simeq 1\dgr.3$.
The cluster is somewhat elongated in the east--west direction, in coincidence with the western large scale structure (LSS) filament \citep{WestEtAl95} that connects it with the cluster A1367. There is X-ray \citep{SimionescuEtAl13, UchidaEtAl16}, optical, weak lensing \citep{OkabeEtAl10,OkabeEtAl14}, radio \citep{BrownRudnick11}, and SZ \citep{PlanckComa12} evidence that the cluster is accreting clumpy matter and experiencing weak shocks towards the filament well within the virial radius, around $\theta\sim0.5\dgr$.

The \gama-ray signal in Coma was discussed in {\Coma} and in {\ComaB}.
It is best described as an elongated, elliptical ring, with semiminor axis coincident with the virial radius, oriented toward the western LSS filament; the best fit was obtained for a ratio $\zeta\equiv a/b\simeq 2.5$ of semimajor axis $a$ to semiminor axis $b\simeq 2.1R_{500}$.
A soft X-ray signature, consistent with a leptonic virial shock signal emitted by lower energy CREs, was identified in the low (R1 and R2) bands of the ROSAT all-sky survey \citep[RASS;][]{SnowdenEtAl97}.
The morphologies of the LAT and ROSAT signals are best fit by the same parameters of the VERITAS signal, and are in better agreement with the planar, rather than the shell, model, as anticipated from the planar distribution of LSS around Coma.
The intensities of the VERITAS, LAT, and ROSAT signals agree with each other, within systematic uncertainties, for an approximately flat CRE spectrum.
Interpreting the signal as a virial shock would imply $\xi_e\dot{m}\simeq0.5\%$, to within a systematic uncertainty factor of a few.

\subsection{Coma: SZ}

The radial, azimuthally-averaged and binned profile of the $y$ parameter in Coma is shown in Figure \ref{fig:ComaFit1}.
The measured (error bars) and modelled (curves) profiles are shown both in the full radial range with gNFW--based models (left panel), and zoomed in on the putative shock region with different $\beta$ model variants (right panel).
A flattening and a possible rise in $y(\tau)$ are observed around $\tau\sim2.1$; this was also seen in an analysis of a southwest sector, where it was putatively associated with a weak relic shock \citep{ErlerEtAl15}.
Beyond $\tau\sim2.3$, the profile is seen to fall off, indicative of the presence of the virial shock, as shown below.

Consider first spherically symmetric gas models in the absence of a shock.
We compute the corresponding $y(\tau)$ profiles by integrating the $P_0$ pressure models (Eqs. \ref{eq:gNFW} or \ref{eq:beta}), convolved the resulting map with a $7\arcmin$ FWHM filter, and radially binning it, as described in \S\ref{sec:MethodSZ}.
The resulting profiles (red dotted curves), found by maximizing the likelihood, depend somewhat on the underlying model and on the radial region being examined, as seen by comparing the two panels.

Next, we incorporate a shock, finding a marked improvement in the fit for all three gNFW-based shock models described in \S\ref{sec:MethodSZ} and shown in the left panel: an ICM shock ($3.7\sigma$ for Eq.~\ref{eq:ICMShock}), a virial shock ($3.7\sigma$ for Eq.~\ref{eq:VirialShock}), and an infinitely strong shock ($4.1\sigma$ for $q\to\infty$).
In all three shock models, the (projected, normalized) shock radius is found to be $\tau_s=2.46\pm0.04$.
The best-fit pressure jumps are large, giving high Mach numbers $\Upsilon\sim 75$ for an ICM shock and $\Upsilon\sim 76$ for the virial shock, but the uncertainties here are substantial; a lower limit $\Upsilon>10$ ($\Upsilon>2.5$) can be placed on the latter only at the $1\sigma$ ($2\sigma$) confidence level.

Similar results are obtained for shock profiles based on the $\beta$ model and when considering other radial ranges, as illustrated in the right panel.
Here, a shock is identified at a higher, $6.3\sigma$ confidence level, with a nearly identical $\tau_s=2.48\pm0.07$.
The narrower radial range upstream considered here allows for a stronger upstream ICM component, and thus a weaker shock; for the $0<\tau<3$ range used to produce the right panel we obtain a lower limit $\Upsilon>1.6$ ($\Upsilon>1.4$) at the $1\sigma$ ($2\sigma$) confidence level.

Next, we consider the possible presence of an additional inner ICM shock (subscript $i$) inward of the virial shock (subscript $s$), by generalizing Eq.~(\ref{eq:ICMShock}) according to
\begin{equation} \label{eq:DoubleShock}
P(r) =
\begin{cases}
P_0(r) & r<r_i \, ; \\
q_i^{-1}P_0(r) & r_i<r<r_s \, ; \\
(q_i q_s)^{-1}P_0(r_s)(r/r_s)^{-5/2} & r>r_s \, .
\end{cases}
\end{equation}
where $r_{i,s}$ and $q_{i,s}$ are the radii and pressure jumps of the two shocks; an analogous generalization is applied to Eq.~(\ref{eq:VirialShock}).
The best fit for the putative, weak, inner shock gives a radius $\tau_i=1.6\pm0.2$ and a Mach number $\Upsilon_i=1.1_{-0.1}^{+0.5}$.
However, this shock is not detected at a significant level in the present, azimuthally averaged analysis.
The parameters of the virial shock are not appreciably changed by incorporating the weak shock.

The radial flattening and possible rise around $\tau\simeq 2.1$ may suggest some underlying structure or morphological change.
This could, in particular, be associated with the elongated leptonic signatures ({\Coma}, {\ComaB}) and with evidence for non-sphericity in published SZ maps \citep{PlanckComa12, KhatriGaspari16} of Coma.
We therefore consider models for simple deviations from sphericity, as illustrated in the right panel.
A sharp transition from spherical to prolate at some radius $\tau_b$ (assuming $\zeta=2.5$ and a constant pressure between a sphere of radius $\tau_b$ and a spheroidal with semiminor axis $\tau_b$) gives a slightly ($0.5\sigma$) better fit with $\tau_b\sim 2.3$.
A sharp transition from spherical to filamentary at some $\tau_f$ gives a noticeably better fit ($2.7\sigma$) for $\tau_f\sim 1.9$.
A more detailed analysis is deferred to future work.

To test if the identification of the virial shock and its parameters are sensitive to the assumed model and its applicability at small radii, we repeat the analysis but restrict it to large, $r>R_{500}$ radii only.
The results are consistent with the full-range analysis, both in best-fit values and in confidence levels.
We also examine SZ maps of the \emph{Planck} collaboration \citep{PlanckXXII16}, extracted with the MILCA algorithm and binned by \citet[][figure 2 therein; see discussion in \S\ref{sec:Discussion}]{KhatriGaspari16}; the results do not appreciably change.

\section{Abell 2319}
\label{sec:A2319}

Abell 2319 is the cluster with the highest signal-to-noise detection in the \emph{Planck} SZ catalogs \citep[$\mbox{SNR}\sim50$;][]{PlanckXXII16}.
Here, $M_{500}\simeq 5.8\times 10^{14}M_\odot$, $R_{500}\simeq 1250\kpc$, $\theta_{500}\simeq0\dgr.3205$, and $z\simeq 0.0557$ \citep{PiffarettiEtAl11}.
Although a major merger was reported on small scales \citep{OHaraEtAl04}, the cluster appears quite spherical in SZ (\SZA), so we analyze it as such.
It is interesting to note that a sharp drop in thermal X-ray emission can be seen around $r\sim 3\Mpc$ \citep[figure 2 in][]{GhirardiniEtAl17}, possibly associated with the virial shock we discuss below.

\subsection{A2319: SZ}

\begin{figure}
    \includegraphics[trim={0 1.4cm 2.0cm 0}, clip ,width=9cm]{\myfig{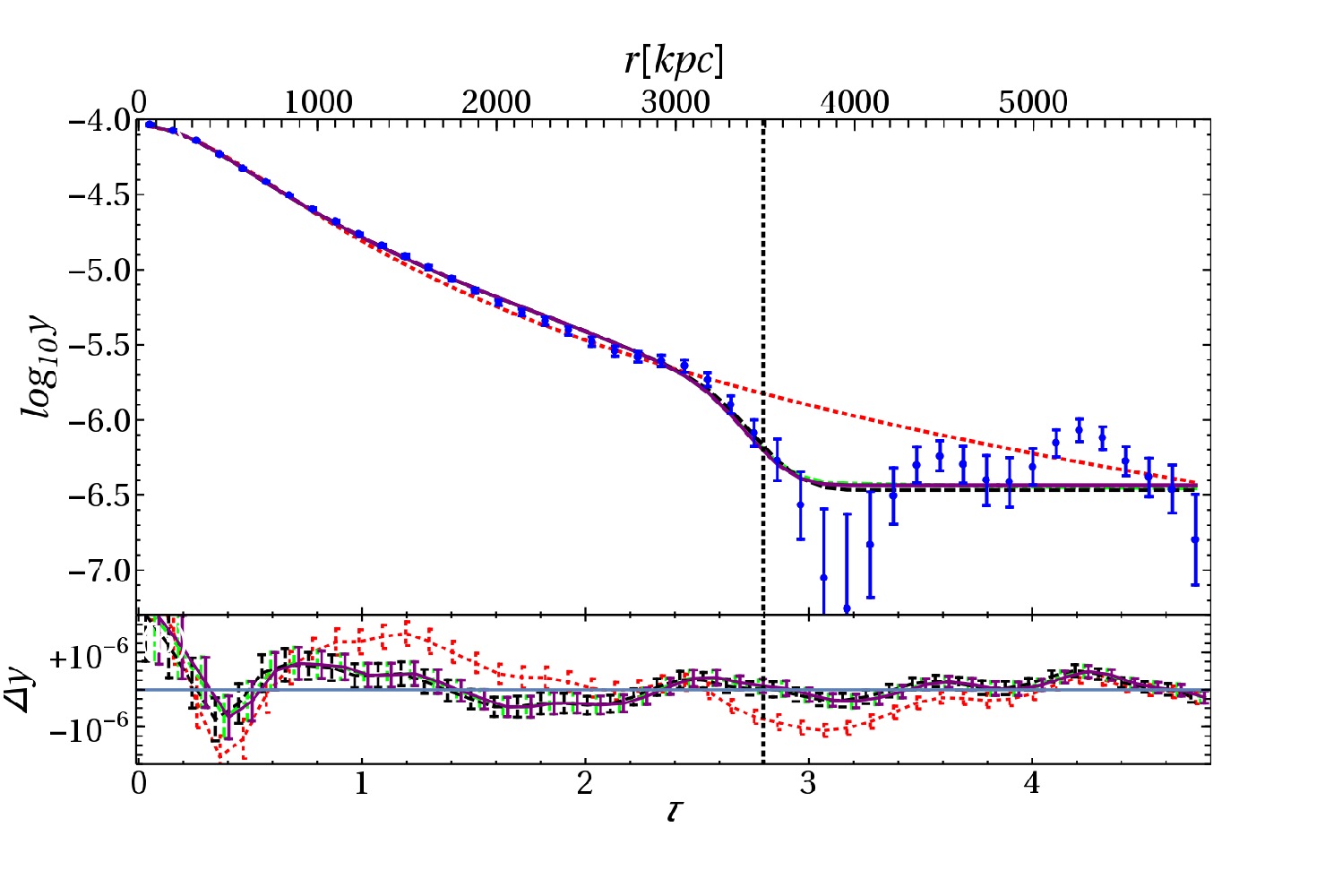}}
    \caption{\label{fig:A2319Fit1}
    SZ y-parameter in A2319 with $\beta$-model based profiles. (For an analogous figure based on the gNFW model, see \SZA.)
    Curves and notations are as in the left panel of Figure \ref{fig:ComaFit1}.
    }
\end{figure}

The radial, azimuthally averaged profile of the $y$-parameter in A2319 was studied by \citet{GhirardiniEtAl17}.
The profile, extracted as detailed in \S\ref{sec:MethodSZ}, is shown in Figure \ref{fig:A2319Fit1}.
The slope becomes steeper around $\tau\sim2.5$, and subsequently flattens beyond $\tau\sim 2.8$.
Note the slight flattening around $\tau\sim2.3$, just before the steepening; this is somewhat reminiscent of the more substantial flattening seen in Coma.

The profile was analyzed by {\SZA} using the gNFW model, and found to harbor a virial shock at the $8.6\sigma$ confidence level.
This shock was identified at a radius $\tau=2.81\pm0.05$ (using the value of $\theta_{500}$ adopted above) and was found to be very strong, with a lower limit $\Upsilon>3.25$ (at the $2\sigma$ confidence level) on the Mach number.
Here, we carry out a complementary analysis based on the $\beta$ model.

We fit the profile by projecting, convolving, and binning the isothermal $\beta$ model variants: without a shock (dotted red curve; Eq. \ref{eq:beta}), with an ICM (dot-dashed green; Eq.~\ref{eq:ICMShock}) or a virial (dashed black; Eq.~\ref{eq:VirialShock}) shock of finite Mach number, and with an arbitrarily strong shock (purple solid; in the $q\to\infty$ limit).
As the shock is inferred to be strong, the different shock curves nearly overlap.

All models give a shock radius consistent with $\tau_s=2.82\pm0.05$, consistent also with the {\SZA} result.
The shock is again found to be strong, with a Mach lower limit $\Upsilon>10$ ($\Upsilon>1.6$) at the $1\sigma$ ($2\sigma$) confidence level.
The detection confidence level is very high, reaching $14\sigma$ for the case of an asymptotically strong shock.
This is higher than found with the gNFW model in {\SZA}, due to the simpler model and the wider radial extent taken into account.

To test if the shock detection is sensitive to the model and to its applicability at small radii, we repeat the analysis but restrict it to large, $r>R_{500}$ radii only, as in the Coma analysis. Here too, the results do not significantly change with the when considering only large radii.

\subsection{A2319: $\gamma$-rays}

\begin{figure}
	\centerline{\epsfxsize=8.cm \epsfbox{\myfig{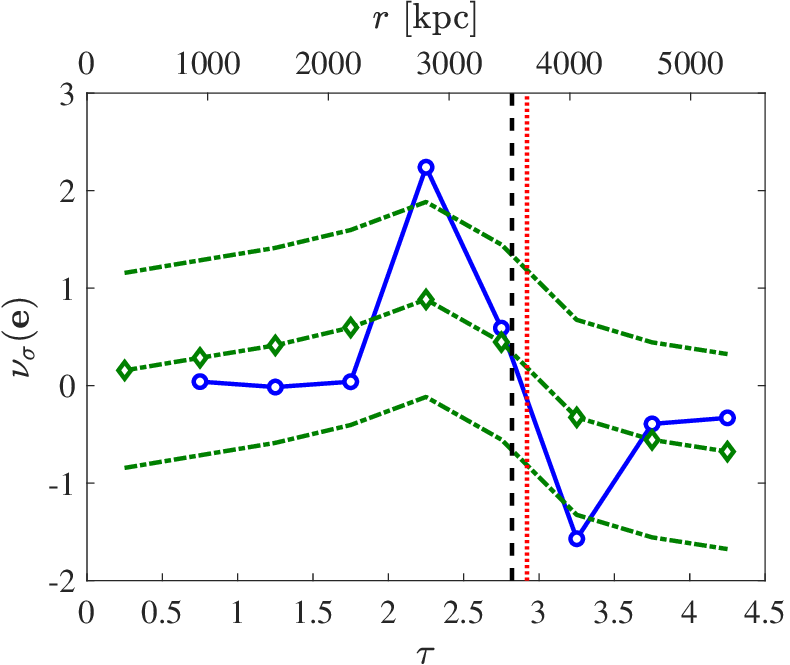}}}
\caption{\label{fig:A2319_significance}
	  The co-added significance of Eq.~(\ref{eq:SingleBinSignificance}) in A2319 is shown for bins of width $\Delta\tau=0.5$ (circles, with solid blue line to guide the eye).
	Also shown are the results (with best fit parameters) of the ring model (diamonds, with green dot-dashed curves to guide the eye and to show the $1\sigma$ intervals).
	The best-fit position of the \gama-ray (SZ) virial shock is shown by a red dotted (black dashed) vertical line.
}
\end{figure}

The A2319 cluster lies near the Galactic plane, at a latitude $\bar{b}\simeq 13.5\dgr$.
Due to the strong $\gamma$-ray Galactic foreground at such low latitudes, here we adopt a fixed (best-fit constant) foreground, and limit the analysis to the close vicinity of the cluster. A nearby point source (3FGL J1913.9+4441) further limits the available analysis area, so we use only a $90\%$, rather than $95\%$, PSF masking.
The significance of the \gama-ray excess emission is presented in Figure \ref{fig:A2319_significance}.

An excess of $\sim2.2\sigma$ can be seen in Figure \ref{fig:A2319_significance} in the $2<\tau<2.5$ bin.
This is the same bin that showed a strong excess in the stacked LAT signal of {\Stack}.
Note that the mock clusters in the control samples of {\Stack} show no such feature.

We model the signal using the spherical accretion shock model of {\Stack} and adopting the $\beta$-model parameters of \citet{FukazawaEtAl04}, in order to translate the emitted \gama-ray flux to electron acceleration rate $\xi_e\dot{m}$.
The inferred shock radius is $\tau_s=2.9^{+0.3}_{-0.4}$, somewhat larger than in the stacked clusters of \Stack, but consistent with the SZ result for this cluster (illustrated by the proximity of the vertical lines in Figure \ref{fig:A2319_significance}).

The inferred acceleration rate is $\xi_e \dot{m}=(0.4\pm0.2)\%$, consistent with previous estimates in other clusters ({\Coma}, {\Stack}, and {\ComaB}).
The TS-based significance (omitting the innermost, $\tau<0.5$ bin, which is adversely affected by point source masking) is low, $1.2\sigma$, but this is mainly driven by the spectral dependence, which may be contaminated at such low latitudes.
One may adopt the mean acceleration efficiency inferred from the stacking of LAT clusters, $\xi_e \dot{m}\simeq 0.6\%$, as a prior for the \gama-ray analysis.
This slightly raises the significance of shock detection to $1.6\sigma$, giving $\tau_s=3.0\pm0.3$.

The overlap between SZ and \gama-ray estimates for the shock radius supports the viability of the \gama-ray signal.
We may use the SZ result, which tightly constrains the shock radius as $\tau_s=2.82\pm0.05$, as a prior for the \gama-ray analysis.
This slightly raises the significance of shock detection in LAT \gama-rays to $1.7\sigma$, leaving the acceleration rate estimate $\xi_e \dot{m}=(0.4\pm0.2)\%$ unchanged.
As expected, a joint SZ--\gama-ray analysis yields a combined shock detection at a very high ($>9\sigma$) confidence level, due to the high significance of the SZ signal.

As a consistency check, we examine a planar leptonic model (as invoked for Coma in \ComaB), in which the shock is assumed to be confined to the plane of the sky. As expected, such a planar model does not provide a better fit for A2319.

\section{Abell 2142}
\label{sec:A2142}

Abell 2142 is the largest and most massive of the three clusters, with $M_{500}\simeq8.15\times 10^{14}M_\odot$, $r_{500}\simeq 1380\kpc$, $\theta_{500}\simeq 0\dgr.2297$.
At a redshift $z\simeq 0.0894$, the cluster shows a very dense core and substantial surrounding substructure \citep{Einasto15}.
The cluster appears quite spherical in SZ, so we analyze it as such.

\subsection{A2142: SZ}

\begin{figure}
    \includegraphics[trim={0 1.4cm 2.0cm 0}, clip ,width=9cm]{\myfig{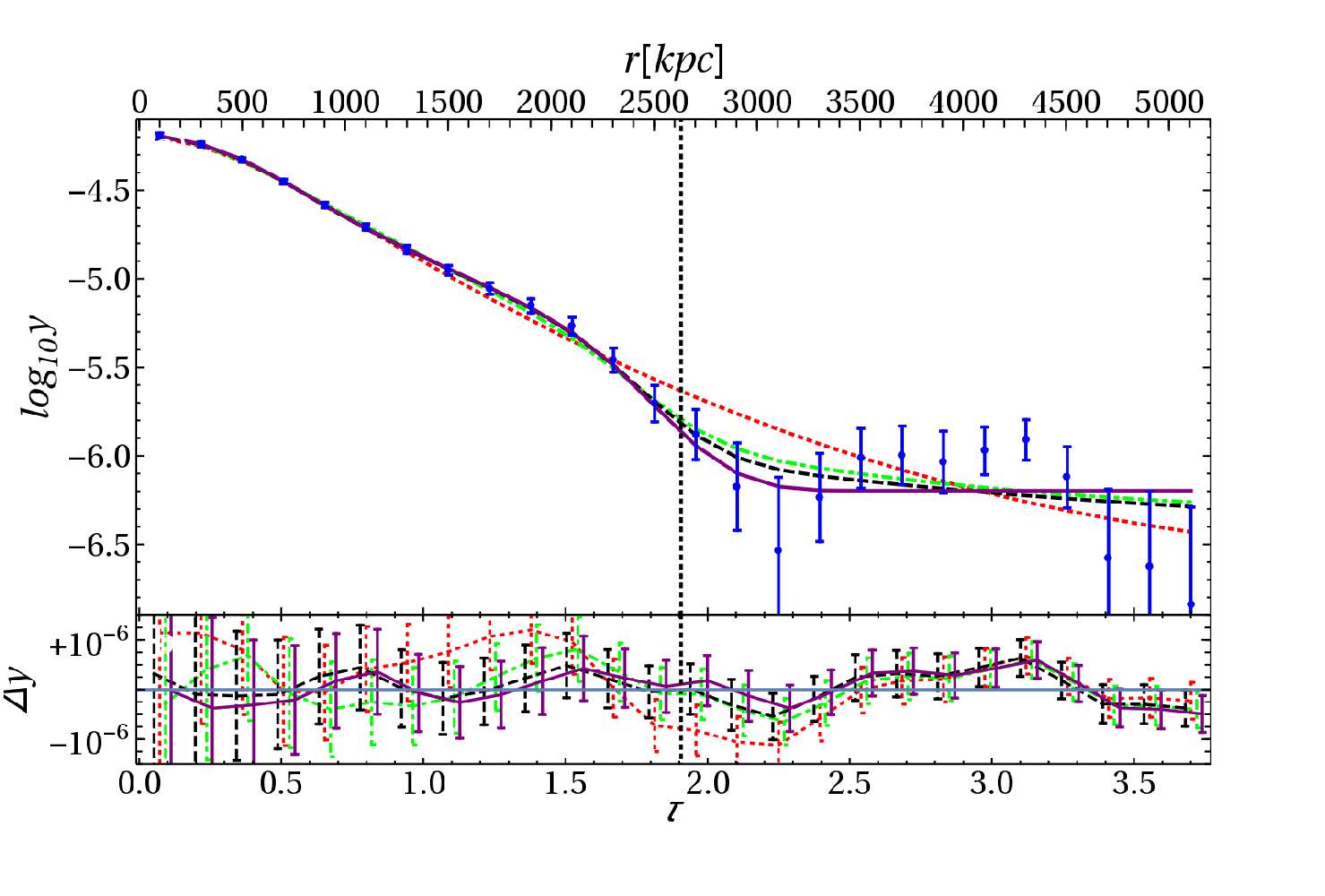}}
    \caption{\label{fig:A2142Fit1}
    SZ y-parameter in A2142 with gNFW based models.
    Curves and notations are as in the left panel of Figure \ref{fig:ComaFit1}.
    }
\end{figure}

The radial, azimuthally averaged profile of the $y$-parameter in A2142, extracted as detailed in \S\ref{sec:MethodSZ}, is shown in Figure \ref{fig:A2142Fit1}.
A steepening can be seen around $\tau\sim1.6$, flattening beyond $\tau\sim2.2$.
Here, the flattening seen just inward of the steepening (compare the data points to the non-shock, dotted-red curve) is very modest.

We first fit the profile with the gNFW-based models.
Adding a shock provides a better fit to the data in all three shock variants.
A strong virial shock is detected at the $3.1\sigma$ confidence level.
All three shock models are consistent with a shock radius $\tau_s=1.89\pm0.06$.
Beyond this radius, some upstream component may still be included in the fit, so the Mach number in principle does not have to be very high; the inferred lower limit is $\Upsilon>2.2$ ($\Upsilon>1.9$) at the $1\sigma$ ($2\sigma$) confidence level.

We test if the shock detection is sensitive to the model and to its applicability at small radii.
First, we repeat the analysis, using the $\beta$ model instead of the gNFW model; the results do not change significantly.
Next, we restrict the analysis to large, $r>R_{500}$ radii only; again, the results remain similar.

\subsection{A2142: \gama-rays}

\begin{figure}
	\centerline{\epsfxsize=8.cm \epsfbox{\myfig{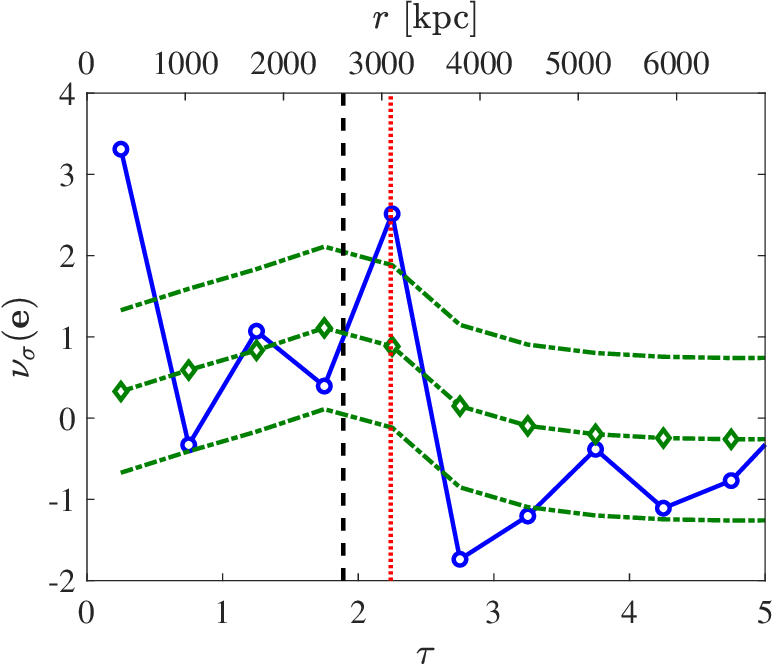}}}
\caption{\label{fig:A2142_significance}
    Same as Figure \ref{fig:A2319_significance}, but for A2142.
	}
\end{figure}

As A2142 lies at an intermediate latitude ($\bar{b}\simeq 48.7\dgr$), here we adopt our nominal, fourth-order polynomial as in {\Stack}, with a $95\%$ point-source masking.
The analysis is otherwise identical to that of A2319.
The significance of the \gama-ray excess emission is shown and modeled in Figure \ref{fig:A2142_significance}.

Again, an excess of $\sim2.2\sigma$ can be seen in the $2<\tau<2.5$ bin --- the same bin showing excess emission in A2319 and in the stacked clusters in {\Stack}.
The TS-based significance of the excess is $2.2\sigma$ (omitting the innermost, $\tau<0.5$ bin due to possible contamination by one high energy photon; see below).
The inferred shock radius is $\tau_s=2.2^{+0.2}_{-0.3}$, and the acceleration rate is $\xi_e \dot{m}=(0.7\pm0.3)\%$, consistent with, and marginally higher than, estimated in other clusters.
We may again use the mean value $\xi_e\dot{m}\simeq 0.6\%$ inferred from the stacking of LAT clusters as a prior for the analysis, yielding a $2.6\sigma$ detection of a shock at $\tau_s\simeq 2.2$.

The $\tau_s$ estimates from SZ and from \gama-rays are again consistent with each other within $1\sigma$.
Using the tightly constrained value $\tau=1.89\pm0.06$ from the SZ profile as a prior for the \gama-ray analysis marginally raises the \gama-ray significance to $2.3\sigma$; the corresponding acceleration rate is $\xi_e \dot{m}=(0.60_{-0.25}^{+0.36})\%$.
A joint SZ and \gama-ray analysis yields a virial shock signal at the $3.5\sigma$ confidence level.
If we fix $\xi_e \dot{m}=0.6\%$ to the stacking value, the joint confidence increases to $3.9\sigma$.

As expected, replacing the shell model with the planar model does not improve the fit for A2142.
The $\sim3.3\sigma$ excess evident in Figure \ref{fig:A2142_significance} at the innermost, $\tau<0.5$ bin is almost entirely due to the single photon detected in the highest energy band; its significance is inflated due to the low foreground expected at such a small bin.

\section{Summary and Discussion}
\label{sec:Discussion}

\MyApJ{\begin{table*}}
\MyMNRAS{\begin{table*}}
\begin{center}
\caption{\label{tab:Summary} Cluster and virial shock parameters in the SZ and \gama-ray analyses.
}
\begin{tabular}{lccc}
\hline
Parameter & Coma & A2319 & A2142 \\
%\MyApJ{\tablenotemark{a} \\}
\hline
Redshift $z$ & $0.0231$ & $0.0577$ & $0.0894$ \\
Radius $r_{500}$ [Mpc] & $1.14$ & $1.25$ & $1.38$ \\
Mass $M_{500}\,[10^{14}M_{\odot}]$ & $4.28$ & $5.83$ & $8.15$ \\
SZ scaled shock radius $\tau_s$ & \mybm{$2.46\pm0.04$} & \mybm{$2.82\pm0.05$} & \mybm{$1.89\pm0.06$}  \\
SZ shock significance (TS)  & $4.1\sigma$ & $14\sigma$ & $3.1\sigma$ \\
Mach number limit ($1\sigma$) & $10$ & $10$ & $2.2$ \\
Mach number limit ($2\sigma$) & $2.5$ & $1.6$ & $1.9$ \\
LAT excess significance & $3.4\sigma$ & $2.2\sigma$ & $2.2\sigma$ \\
LAT shock significance (TS) & $2.5\sigma$ & $1.2\sigma$ & $2.2\sigma$ \\
LAT constrained significance (TS) & --- & $1.7\sigma$ & $3.9\sigma$ \\
LAT shock radius $\tau$ & \mybm{$>\tau_b=2.14_{-0.06}^{+0.07}$} & \mybm{$2.9_{-0.4}^{+0.3}$} & \mybm{$2.2_{-0.3}^{+0.2}$} \\
Acceleration rate $\xi_e\dot{m}\,[1\%]$ & $0.19\pm0.07$ & $0.4\pm0.2$ & $0.7\pm0.3$ \\
%\vspace{0.5mm}
\hline
\end{tabular}
%\MyApJ{\tablenotetext{1}{Comment if needed.}}
\end{center}
\vspace{0.6cm}
\MyApJ{\end{table*}}
\MyMNRAS{\end{table*}}

Motivated by the detection of leptonic signals from the virial shocks of galaxy clusters (\Coma, \Stack, and \ComaB) and by preliminary evidence for a morphological coincidence between this leptonic emission and an SZ cutoff on the $y$-parameter (\Coma), we present a joint analysis of SZ data from \emph{Planck} (Figures \ref{fig:ComaFit1}, \ref{fig:A2319Fit1}, and \ref{fig:A2142Fit1}) and of \gama-ray data from the LAT (Figures \ref{fig:A2319_significance} and \ref{fig:A2142_significance}).
The analysis focuses on Coma (Figure \ref{fig:ComaFit1}) and A2319 (Figures \ref{fig:A2319Fit1}--\ref{fig:A2319_significance}), for which \gama-ray or SZ virial signals were already published, and supplements them with a new joint analysis of \gama-ray and SZ data in A2142 (Figures \ref{fig:A2142Fit1}--\ref{fig:A2142_significance}), selected by its high mass and by data availability.
Our results are summarized in Table \ref{tab:Summary}.

The imprint of the virial shock on the radial profile of the SZ $y$-parameter is detected at fairly high confidence levels in all three clusters, reaching $8.6\sigma$ in A2319 (\SZA), $4.1\sigma$ in Coma, and $3.1\sigma$ in A2142.
These confidence levels are obtained with gNFW-based models; comparable or higher confidence levels are found when using the simple $\beta$ models, instead.
When incorporating the virial shock in either gNFW or $\beta$ models, they fit the data well in all cases.
Consistent shock parameters are inferred from the different model variants, even when masking small or large radii.

The LAT \gama-ray excess near the virial radius is found to be $2.2\sigma$, or slightly higher, in all three clusters.
Such confidence levels are to be expected, based on previous studies (\Stack, \ComaB), when analysing an individual cluster, due to the limited photon statistics, the instrumental point spread function, and the strong Galactic foreground.
The TS-based statistics show comparable significance levels, except in A2319, which suffers from a considerable Galactic foreground due to its low latitude.
Adopting the mean CRE acceleration rate inferred from a previous stacking of LAT clusters (\Stack) reduces the fit to a one-parameter model, raising the significance of shock detection in all cases.

The detection of the \gama-ray and SZ signals, and their interpretation as associated with each other and arising from the virial shock, are further supported by the inferred radii and properties of the virial shocks.
First, while the shock radius $\tau_s$ is treated as a free parameter in each analysis of each cluster, it is found to be near the virial radius in all cases.
Second, while in each cluster $\tau_s$ is treated independently in \gama-ray and SZ analyses, the two values are found to be consistent with each other, within $1\sigma$, in A2319 and A2142 (illustrated by the proximity of the vertical lines in Figures \ref{fig:A2319_significance} and \ref{fig:A2142_significance}); in Coma, non-sphericity complicates the comparison.
Third, the shock Mach numbers are found to be high, as expected in a virial shock.
And fourth, the acceleration rates $\xi_e\dot{m}$ inferred from the \gama-ray signals are consistent among the three clusters (see discussion below) and with previous studies.

The SZ signature of a virial shock can be used to boost the sensitivity for the detection of the leptonic emission from the shock, or vice versa.
One option is to use the shock radius inferred from SZ as a prior for the \gama-ray analysis.
Another option is to carry out a joint SZ--\gama-ray analysis.
Both methods are shown to raise the significance of the \gama-ray detection.
For example, a joint analysis of A2142 yields a $3.5\sigma$ shock detection.
Further fixing the acceleration rate on the value inferred from stacking other clusters yields a higher, $3.9\sigma$ signal.

The SZ signals upstream of the shock are found to be near or below the background.
Consequently, the inferred shock parameters are insensitive to assumptions on the upstream plasma.
Lower limits can still be imposed on the strength of the virial shock.
However, due to the uncertain foreground level and the low signal-to-noise near and beyond the virial radius, we are only able to impose lower limits with substantial uncertainty; the $2\sigma$ upper limits are of order $\Upsilon\sim 2$.

The acceleration rates $\xi_e\dot{m}$ inferred in the three clusters support previous estimates of order a few $0.1\%$.
Modeling the SZ signal and the distributions of galaxies near the virial shock, one can measure $\dot{m}$ and break its degeneracy with $\xi_e$.
For example, in A2319, the accretion rate was estimated (\SZA) as $\dot{m}_{200}\simeq 2.0$.
Adopting the $\beta$ model scalings (\Stack) $\dot{m}\propto \delta^{1/2}\propto r_\delta^{-1}$, the accretion rate at the A2319 shock is found to be $\dot{m}_{63}\simeq 1.1$, which gives $\xi_e\simeq 0.5\%$ in this cluster.

We find that the parameters of both gNFW and $\beta$ models change substantially when taking into account the presence of a shock.
This suggests that when modeling data beyond $\sim1.5R_{500}$, the projected effect of the virial shock must be taken into account.

Some flattening of the $y$-parameter profile is seen just inward of the virial shock.
This is most pronounced in Coma, but is also seen in A2319, and possibly also in A2142.
The flattening may be associated with recently accreted substructure or with a change in morphology, as suggested by the improved fit obtained for Coma when adding a filamentary component.

We find that the presence and parameters of the virial shocks can be inferred from the SZ data using the $\beta$ model, without invoking the more complicated gNFW model; the results of the two approaches are consistent with each other.
We similarly examine if the error bars $\sigma_d(y_j)$, representing the diagonal of the covariance matrix, can be used to give a reasonable estimate of the uncertainties even without accounting for the off-diagonal terms.
We find that for the parameters used in this study, the correlations among neighboring bins in the radial $y$ parameter profile can be approximately accounted for by co-adding to each diagonal term the mean differences in $y$ between the neighboring bins, $\sigma(y_j)^2=\sigma_d(y_j)^2+(y_j-y_{j+1})^2/2+(y_j-y_{j-1})^2/2$.
This is analogous to accounting for an unknown position within the radial bin.
In this approximate method, we can analyze previously published $y$ parameter profiles \citep{KhatriGaspari16, GhirardiniEtAl17}, recovering the same results obtained here from the full analysis.

We note that the acceleration rates $\xi_e\dot{m}$ inferred in the three cluster seemingly show a mild correlation with the cluster mass, $M_{500}$.
However, this tendency is not significant, and was not found among the 112 clusters analyzed by {\Stack}.
Furthermore, the rate inferred in Coma pertains to a different, prolate model, so there is a considerable systematic uncertainty when comparing it to the other clusters.

\MyApJ{\acknowledgements}
\MyMNRAS{\section*{Acknowledgements}}
We thank C. Tchernin and A. Zitrin for helpful discussions.
This research was supported by the Israel Science Foundation (ISF grant No. 1769/15),
by the IAEC-UPBC joint research foundation (grant No. 300/18),
and by the GIF (grant I-1362-303.7/2016).
G.H. acknowledge support from Spanish Ministerio de Econom\'ia and Competitividad (MINECO) through grant number AYA2015-66211-C2-2.

\bibliography{Virial}

\end{document}